\documentclass[prb,aps,showpacs,amsmath,amssymb,superscriptaddress,affilletter,twocolumn]{revtex4-1}
\pdfoutput=1

\usepackage[usenames,dvipsnames]{color}
\usepackage{graphicx}
\usepackage{tabularx}
\usepackage[T1]{fontenc}
\usepackage[cp1250]{inputenc}
\usepackage{dcolumn}
\usepackage{amsmath,amsfonts,amssymb}
\usepackage{multirow}
\usepackage[rightcaption]{sidecap}
\usepackage{float}
\newcommand{\Xm}{X$^-$}
\newcommand{\Xp}{X$^+$}
\newcommand{\spl}{\sigma^+}
\newcommand{\smn}{\sigma^-}
\newcommand{\Xz}{X$^0$}

\newcommand{\CdMnTe}{Cd$_{1-x}$Mn$_x$Te}

\newcommand{\beq}{\begin{equation}}
\newcommand{\eeq}{\end{equation}}

\begin{document}

\title{Optical signatures of spin dependent coupling in semimagnetic quantum dot molecules}

\author{Ł.~Kłopotowski}
\email[Corresponding author: ]{lukasz.klopotowski@ifpan.edu.pl}
\affiliation{Institute of Physics, Polish Academy of Sciences, Al. Lotników 32/46, 02-668 Warsaw, Poland}

\author{P.~Wojnar}          \affiliation{Institute of Physics, Polish Academy of Sciences, Al. Lotników 32/46, 02-668 Warsaw, Poland}

\author{Ł.~Cywiński}          \affiliation{Institute of Physics, Polish Academy of Sciences, Al. Lotników 32/46, 02-668 Warsaw, Poland}

\author{T.~Jakubczyk}            \affiliation{Faculty of Physics, University of Warsaw, ul. Pasteura 5, 02-093 Warsaw, Poland}

\author{M.~Goryca}            \affiliation{Faculty of Physics, University of Warsaw, ul. Pasteura 5, 02-093 Warsaw, Poland}

\author{K.~Fronc}           \affiliation{Institute of Physics, Polish Academy of Sciences, Al. Lotników 32/46, 02-668 Warsaw, Poland}

\author{T.~Wojtowicz}       \affiliation{Institute of Physics, Polish Academy of Sciences, Al. Lotników 32/46, 02-668 Warsaw, Poland}

\author{G.~Karczewski}      \affiliation{Institute of Physics, Polish Academy of Sciences, Al. Lotników 32/46, 02-668 Warsaw, Poland}

\date{\today}

\begin{abstract}
We present photoluminescence studies of CdTe and \CdMnTe\ quantum dots grown in two adjacent layers. We show that when the dots are 8 nm apart, their magnetooptical properties -- Zeeman shifts and transition linewidths -- are analogous to those of individual CdTe or \CdMnTe\ dots. When the dots are grown closer, at a distance of 4 nm, it becomes possible to tune the electron states to resonance and obtain a formation of a molecular state hybridized over the two dots. As a result of the resonant enhancement of the electron-Mn ion exchange interaction, spectroscopic signatures specific to spin-dependent inter-dot coupling appear. Namely, an anomalous increase of the Zeeman shift and a resonant increase in the transition linewidth are observed. A simple model calculation allows us to quantitatively reproduce the experimental results.
\end{abstract}


\maketitle

\section{Introduction}

Coupled semiconductor quantum dots (QDs) constitute an ideal system in which  zero-dimensional quantum states can be efficiently manipulated. In view of this potential, coupled QDs are used for implementation of qubit gates. Indeed, the qubit state in an electrostatically defined pair of QDs has been coherently manipulated and an operation of a quantum gate based on this system has been demonstrated.\cite{pet05,kop06} Unlike the electrostatic dots, self-assembled QDs are optically active \cite{hen08} and thus can be exploited in, e.g., coupling to flying qubits,\cite{ima99} or transmission of quantum information as photons over macroscopic distances.\cite{rau14} Controllable coupling of two self-assembled QDs allows to tailor the carrier wave function and, in particular, to form a molecular-like bonding and antibonding states hybridized over the two dots. Such structures, quantum dot molecules (QDMs), allow for, e.g., spectroscopy of single carrier excited states,\cite{sch08,mue11} exciton storage for a time exceeding its radiative lifetime by three orders of magnitude,\cite{boy11} efficient tuning of the electron {\em g-}factor,\cite{dot09} or the control over anisotropic electron-hole exchange splitting \cite{sko13,bry13} -- the crucial parameter controlling the generation efficiency of entangled photon pairs. From the point of view of applications as quantum gates, two important properties have been demonstrated: optical response of one QD conditional on the quantum state of the other QD \cite{rob08,mue11} and optical coherent manipulation of a two spin qubit state.\cite{kim11}.

All the exciting properties of QDM rely on controlling the tunnel coupling between the two QDs. To date, this is achieved by tuning the energies of single carrier states in the two dots with external electric field.\cite{kre05,bei06,sti06,bra06,sch07prb,dot10} At the resonance, mixing of the orbital states leads to formation of the molecular-like, bonding and anti-bonding states. In a spectroscopic experiment, the splitting between them, proportional to the tunneling rate, gives rise to specific spectral patterns. For neutral excitons, an anticrossing of transition lines is observed.\cite{kre05,bra06} It is related to an anticrossing in the initial state of the recombination process: between a spatially direct state with the electron and hole localized in the same QD and an indirect exciton -- with the electron and hole occupying adjacent dots. For charged excitons and a biexciton, anticrossings in the initial and final states give rise to more complicated X-shape patterns, \cite{sti06,sch07bi} which reveal a fine structure originating from the electron-hole exchange interaction.\cite{sch07prb,dot10}

In this work, we investigate an alternative way of controlling the coupling in a QDM. Namely, we use a magnetic field to tune the single carrier states to resonance. Since the intrinsic {\em g-}factors of InAs or CdTe QDs give rise to a Zeeman splitting on the order of 0.2 meV/T, we employ QDM with a non-magnetic CdTe QD and a semimagnetic \CdMnTe\ QD.\cite{mak00,bac02,hun04,cle10,klo11pol,klo13} In the latter one, the Zeeman splitting can be amplified by up to two orders of magnitude as a result of the exchange interaction between the charge carriers and Mn ions.\cite{gaj10} Since the Zeeman effect lifts the degeneracy between the spin-up and spin-down states, in such a configuration, only states with one spin polarization are brought to resonance. Consequently, the coupling becomes spin dependent as the molecular state is formed from specific spin states. In the photoluminescence (PL) experiment we observe coupling signatures different from those reported for InAs-based QDMs. Our interpretation relies on the analysis of the Zeeman energy shifts and changes of the PL linewidth with magnetic field, which unambiguously shows the formation of a molecular state. 
We remark that, apart from novel coupling schemes, semimagnetic QDMs are predicted to exhibit a very efficient tunability of the {\em g-}factors with electric field used to tailor the wave function overlap with the Mn ions.\cite{lya12} This property should in turn allow for fast qubit rotations utilizing electric fields from optical fields.\cite{pre08}

\section{Samples and Experiment}

The samples are grown by molecular beam epitaxy on a (100)-oriented GaAs substrate. A CdTe buffer, 4 $\mu$m thick is grown first. Then, a ZnTe layer, 700 nm thick, is grown. The QDs are formed on top of this layer employing a modified Stranski-Krastanow mechanism.\cite{tin03,woj10} In order to separate the PL signal from the CdTe and \CdMnTe\ QDs, it is important to obtain a layer of pure CdTe dots without any residual Mn ions. However, as shown extensively for \CdMnTe\ quantum wells, the Mn ions migrate along the growth direction resulting in smearing of the interfaces.\cite{gri96} Therefore, in order to avoid the migration of the Mn ions into the CdTe QDs, a layer of CdTe QDs is grown first. Then, a ZnTe spacer layer is deposited on top of which the second QD layer formed from \CdMnTe, is grown with intentional Mn concentration $x=0.05$. The top QD layer is then covered with a ZnTe barrier, 50 nm thick.

We study two samples with different thicknesses $d$ of the ZnTe spacer layer, i.e., different distances between the top and bottom dots: $d=4$ nm and $d=8$ nm. In order to access single QDMs, gold shadow mask apertures, 500 nm in diameter are produced on top of the sample by spin-casting polystyrene beads, Au evaporation and lift-off.

Photoluminescence (PL) measurements are performed in a split-coil cryostat at bath temperature of 2 K in magnetic fields up to 5 T. The PL signal is excited with a 532 nm solid state laser. The sample is glued on top of a reflective microscope objective, immersed together with the sample. The signal is detected with a CCD camera coupled to a 0.5 monochromator with an overall spectral resolution better than 100 $\mu$eV. The excitation density is kept low enough to avoid heating of the Mn spin system, which results in anomalous behavior of the PL transitions in magnetic field.\cite{cle10} Unless stated otherwise, we detect two circular polarizations by rotating a half-wave plate placed between a quarter-wave plate and a linear polarizer.

\begin{figure*}
  \includegraphics[angle=0,width=\textwidth]{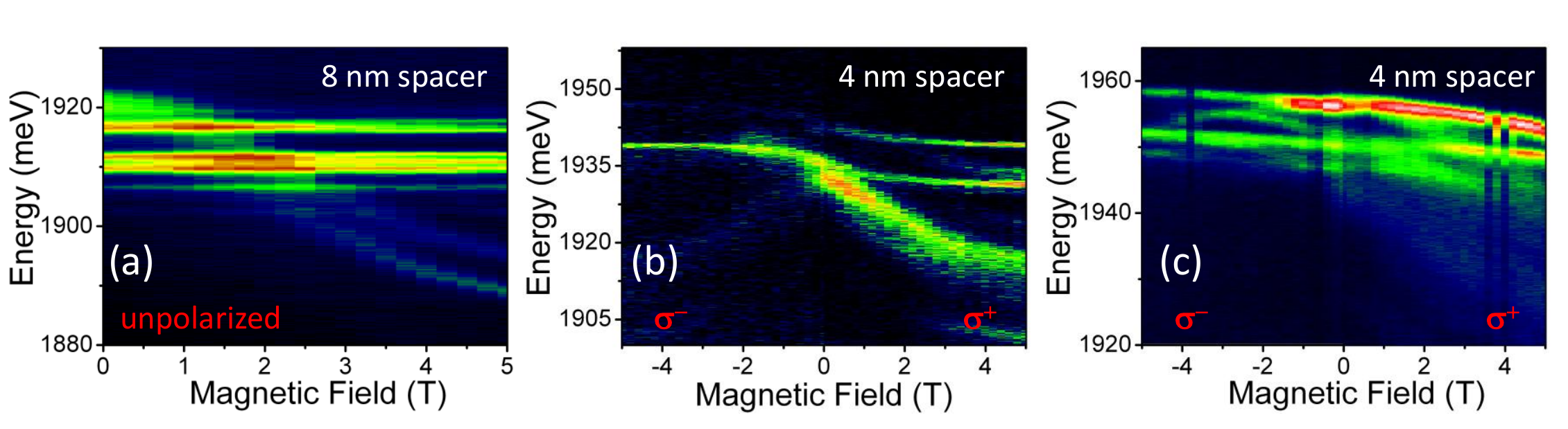}
  \caption{Maps showing magnetic field dependence of the PL for single QDs. (a) Sample with a 8 nm spacer. (b) and (c): PL from two spots on the sample with a 4 nm spacer. Positive and negative magnetic fields denote detection in $\spl$ and $\smn$ polarization, respectively. The spectra in (a) originate from single, uncoupled CdTe nad \CdMnTe\ QDs. The spectra in (b) demonstrate the case of a weak inter-dot coupling, almost independent of the magnetic field. The spectra in (c) show the signature of resonant coupling giving rise to a molecular electron state at about 3T.}  \label{maps}
\end{figure*}

\section{Experimental Results}

We start the discussion of experimental results with the PL of the QDs separated with the 8 nm spacer. In Fig. \ref{maps}(a), we show a map demonstrating the magnetic field dependence of the PL from single QDs. In this measurement, we detected unpolarized light. Two kinds of spectral features are seen: broad transitions exhibiting a giant Zeeman shift toward lower energies and narrow transitions with a small Zeeman splitting -- amounting to about 1.5 meV at 5 T. We interpret these two features as exciton recombinations in the semimagnetic and nonmagnetic QDs, respectively. Both features consist of several lines. In the spectra shown in Fig. \ref{maps}(a), there are two transitions related to the \CdMnTe\ QD and four related to the CdTe dot. Coexistence of these transitions manifest charge state fluctuations occurring on the time scale much shorter than integrations time -- a well established phenomenon in CdTe-based QDs.\cite{suf06,kaz10,klo11,kaz11} The four narrow transitions seen in Fig. \ref{maps}(a) are recombinations of the neutral exciton (\Xz), positively (\Xp) and negatively (\Xm) charged excitons and the biexciton (2X) from the CdTe dot. These transitions appear in a sequence universal for CdTe QDs, namely $E_{X^0} > E_{X^+} > E_{X^-} > E_{2X}$, where $E_{\chi}$ is the emission energy of the complex $\chi$.\cite{klo11} The broad transitions are recombinations from the \CdMnTe\ dot. For these QDs, we usually observe two transitions and increasing the excitation density does not result in an appearance of the biexciton. We interpret this effect as lowering of the valence band confinement upon addition of the manganese. Indeed, as discussed in previous papers\cite{klo11pol,klo14}, the QD potential in the valence band in the CdTe/ZnTe system is mostly due to strain. Increasing the Mn concentration decreases both the lattice mismatch and the bare valence band offset weakening the hole confinement. Eventually, above a certain Mn concentration $x$, the hole-hole Coulomb repulsion precludes charging of the QD with two holes and hence the absence of both the \Xp\ and the 2X in the PL spectra. We therefore interpret the transitions related to the \CdMnTe\ dot as the recombinations of the \Xz\ and \Xm. We remark that the \Xm\ spectroscopic shifts ($\Delta_{X^-} = E_{X^0} - E_{X^-}$) for the semimagnetic QDs are by 20-50\% smaller than the corresponding values for CdTe QDs. For the semimagnetic QD from Fig. \ref{maps}(a), $\Delta_{X^-} \approx 6$ meV, while the average value for CdTe QDs is about 10 meV.\cite{klo11,kaz11} We suppose that this reduction of the spectroscopic shifts reflects decrease of the \Xm\  binding energy resulting from the weaker confinement of the hole in a semimagnetic QD.

The broadening of the transitions originating from the \CdMnTe\ QD are due to the spin state fluctuations of the Mn ions.\cite{bac02} The fluctuations become progressively suppressed with increasing magnetic field and hence the transitions become narrower.\cite{klo13} We note that the upper Zeeman branch for the transitions related to the \CdMnTe\ dot is not seen in Fig. \ref{maps}(a) due to efficient spin and energy relaxation of excitons in these structures.\cite{klo11pol} In anticipation of the results presented below, we remark that the spin relaxation time is strongly dependent on the effective field exerted by Mn on the exciton and thus on the Mn ion density.\cite{klo11pol}

\begin{figure}[!h]
  \includegraphics[angle=0,width=.33\textwidth]{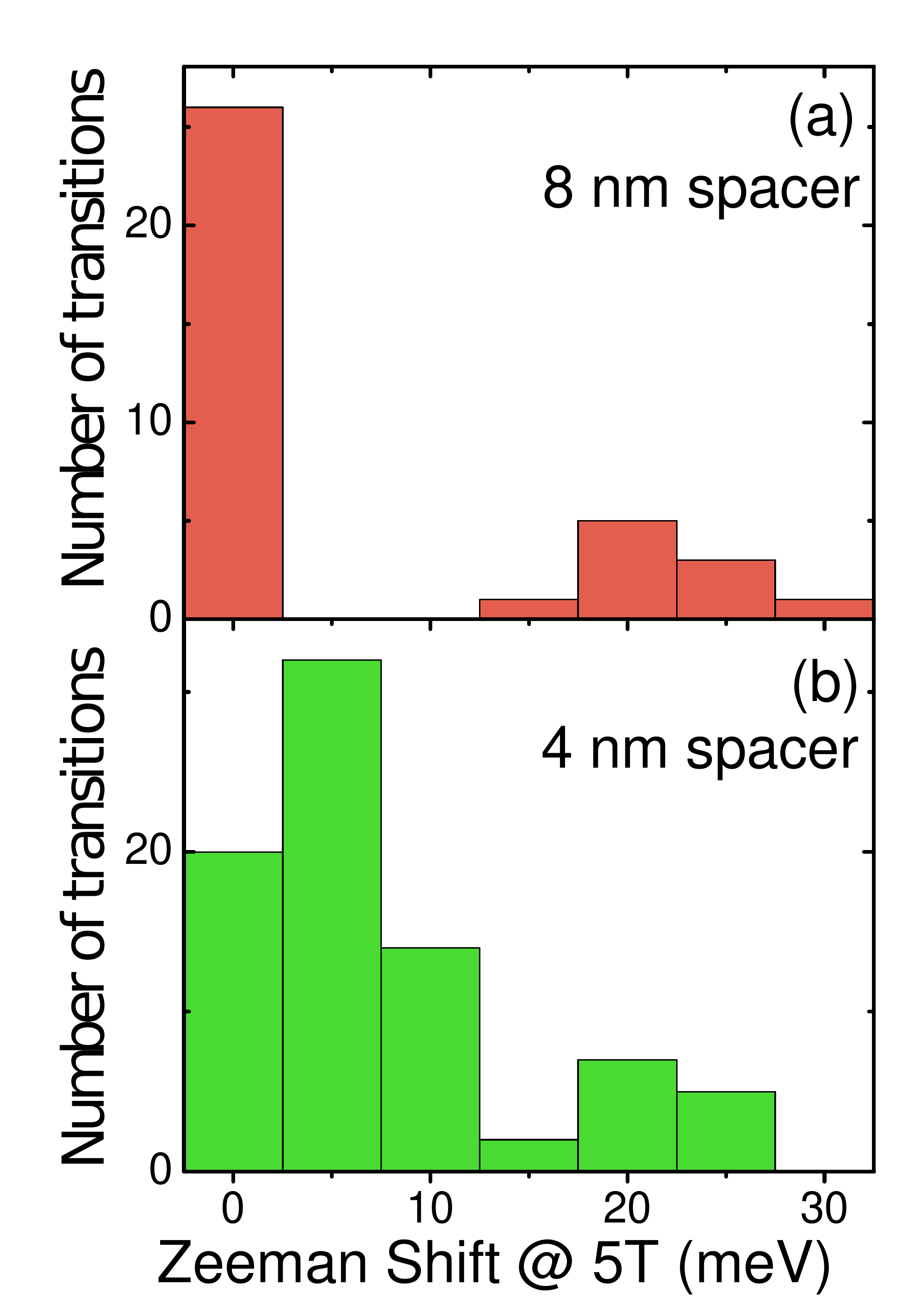}
  \caption{Statistics of the Zeeman shift of single QD PL evaluated at 5 T for the two samples with different widths of the ZnTe spacer layer between the dots. (a) Sample with 8 nm  spacer. (b) Sample with 4 nm spacer.}
  \label{stats}
\end{figure}

While the PL spectra shown in Fig. \ref{maps}(a) can be interpreted as originating from single, {\em uncoupled} CdTe and \CdMnTe\ dot, a qualitatively different results are obtained on the sample, where the QDs are separated with a 4 nm ZnTe spacer. In Figs. \ref{maps}(b) and (c), we present results of polarization resolved PL measurements on two different apertures. Positive and negative magnetic fields denote detection in $\spl$ and $\smn$ polarization, respectively. In both maps, we observe transitions with Zeeman splittings and linewidths significantly larger than for pure CdTe QDs, but smaller than those observed for semimagnetic Cd$_{0.95}$Mn$_{0.05}$Te\ QDs from the sample with 8 nm spacer and single layer samples investigated in Ref. [\cite{klo13}]. This observation is confirmed in Fig. \ref{stats}, where we show Zeeman shift statistics, measured for the $\spl$ detection polarization at 5 T, obtained on several tens of QDs from both samples. For the 8 nm sample the Zeeman shift distribution is clearly bi-modal with one peak below 2 meV and the other one, broader, at about 25 meV. We interpret these peaks as transitions from {\em uncoupled} CdTe and \CdMnTe\ QDs. For the 4 nm sample, these peaks are also present, but the distribution is undeniably different. It is clear that the QD ensemble still contains uncoupled, pure CdTe and \CdMnTe\ dots, but the majority of transitions exhibit intermediate Zeeman shifts. This is a clear indication that the QDs separated by a 4 nm spacer are close enough for the exciton wave functions from one dot to penetrate into the other. As a result, the wave function overlap with the Mn ions varies from dot to dot resulting in a broad distribution of Zeeman shifts. The results presented in Fig. \ref{stats} indicate that inter-dot coupling controlled by the magnetic field is possible in this sample, provided that the available Zeeman splittings are large enough to bring the states in adjacent dots to resonance.

In order to understand the values of the Zeeman shifts and linewidths observed in the PL spectra of the 4 nm sample, we will look more closely at the two cases presented in Figs. \ref{maps}(b) and (c). In the first one, we see a broad feature with a linewidth at 0 T of about 7 meV. The feature comprises of two transitions with an average Zeeman shift of about 18 meV at 5 T, i.e., more than 70\% of the average Zeeman shift observed for the uncoupled \CdMnTe\ dots from the 8 nm sample. These transitions become fully polarized at about 2 T. Apart from that, there are two transitions separated by about 8 meV with linewidths at zero field of about 2.0-2.5 meV. 
The Zeeman shifts at 5 T in this case are about 4 meV, compared to 0.5 meV for pure CdTe QDs from the 8 nm sample (and samples containing single layer of CdTe dots\cite{kaz09}). We identify these narrow transitions as recombinations of the \Xz\ and \Xm. The former becomes almost 100\% polarized by the magnetic field. The polarization degree ($P=(I_+-I_-)/(I_++I_-)$, where $I_{\pm}$ are PL intensities in $\sigma^{\pm}$ polarizations) of the \Xm\ PL reaches only about 50\%. This difference points out that in semimagnetic QDs exciton spin relaxation is faster than spin relaxation of the hole,\cite{klo13} which controls the polarization behavior of the \Xm\ transition.

The question arises, whether the PL behavior presented in Fig. \ref{maps}(b) can be a fingerprint of a controllable coupling in a QDM. The formation of a molecular state occurs when electron or hole states are at resonance and the tunnel coupling results in hybridization of a carrier wave function over the two dots. As the energy of the carrier state in the \CdMnTe\ is tuned with magnetic field with respect to a weakly shifting state in the CdTe dot, on one side of the resonance the ground state is predominantly non-magnetic, while on the other it is predominantly semimagnetic. Therefore, for a our QDM, we should expect a different magnetic field behavior of the PL on both side of the resonance and, in particular, for $\spl$ and $\smn$ transitions. Specifically, since the coupling controls the overlap of the wave function with the Mn ions, and thus the strength of the carrier-ion exchange interaction,
an asymmetry in the field dependence of the Zeeman shifts and linewidths should be expected.

\begin{figure}[!h]
  \includegraphics[angle=0,width=.33\textwidth]{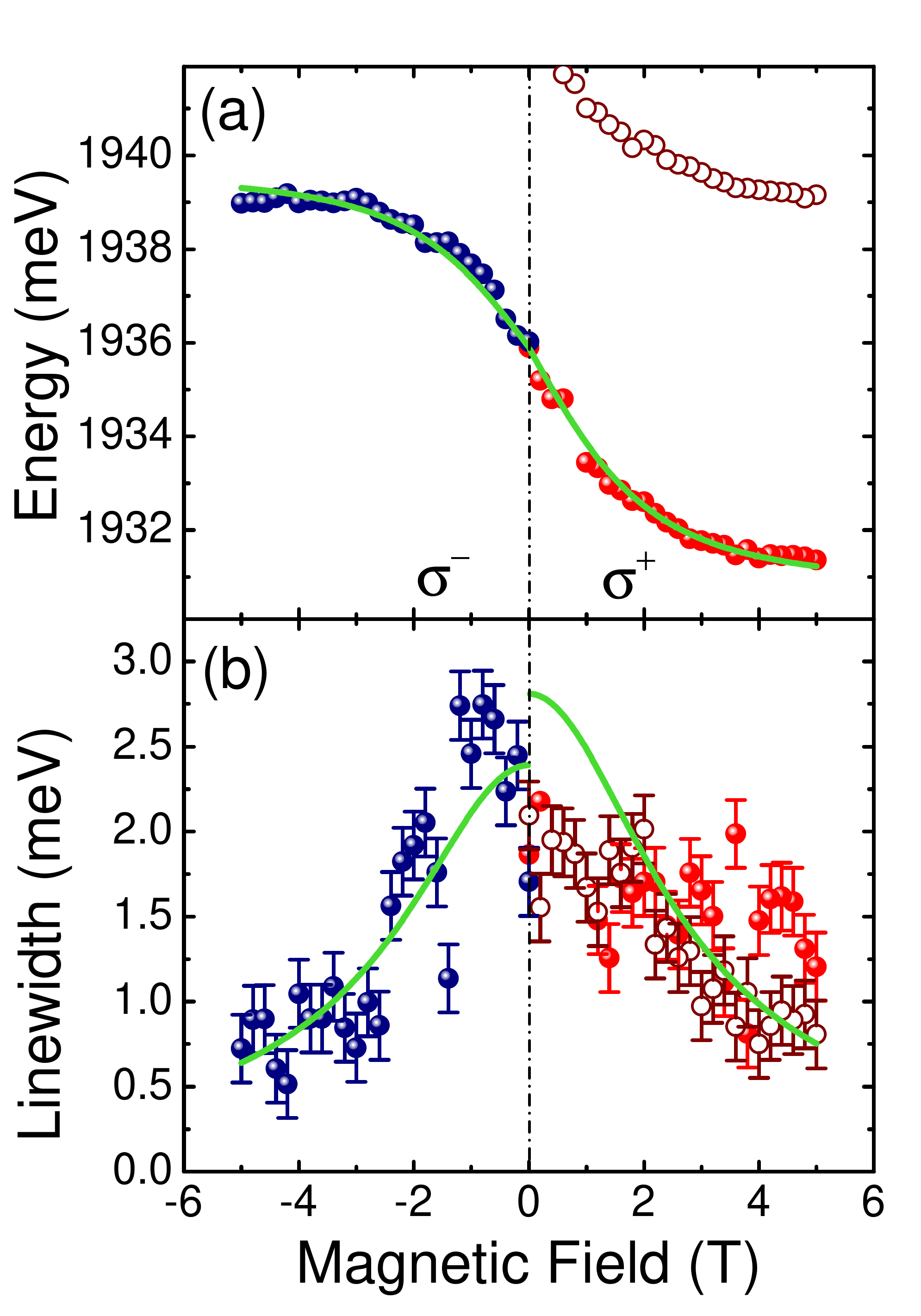}
  \caption{Energy positions (a) and linewidths (b) extracted from the narrow transitions seen in Fig. \ref{maps}(b). Lines are results of fitting of the theoretical model (Eqs. (\ref{eneb}) and (\ref{gammab})). See text for details.}
  \label{unc}
\end{figure}

In Fig. \ref{unc}(a) and (b), we plot, respectively, the transition energies and transition linewidths for the narrow transitions extracted from Fig. \ref{maps}(b). 
Indeed, a small asymmetry in the Zeeman shifts is observed for the \Xm\ transition: the blueshift of the $\smn$-polarized transition is about 3.1 meV, while the $\spl$-polarized transition redshifts by about 4.5 meV. Also, the transition linewidths, although decrease for both polarizations with magnetic field, reach 0.5 meV and 0.8 meV, for the $\spl$- and $\smn$-polarized transitions, respectively.

In order to verify whether the observed asymmetry can be due to formation of a molecular state, we analyze quantitatively the data presented in Fig. \ref{unc}. We extend results of the model presented in Ref. [\cite{klo13}] to describe the transition energies and linewidths resulting from recombination of an exciton with only a partial overlap with the Mn ions in the neighboring QD. We neglect the effect of the intrinsic Zeeman energy in CdTe QDs since it results of Zeeman shifts a factor of 7 smaller than the ones observed in Fig. \ref{maps}(b). Thus, we obtain the following expression for the magnetic field dependence of the transition energies in $\sigma^{\pm}$ polarizations:
\beq
E_{\pm} = E(0) \pm \frac{1}{2} N_0 (\gamma_e \alpha-\gamma_h \beta) x_{\text{eff}} S B_{S}(\frac{g \mu B S}{k_B(T+T_0)}),
\label{eneb}
\eeq
where $E(0)$ is the transition energy at magnetic field $B=0$,$N_0$ is the number of cation sites per unit volume, $N_0 \alpha = 0.22$ eV and $N_0 \beta=-0.88$ eV are, respectively, the exchange integrals for electrons and holes, $g=2$ and $S=5/2$ are the {\em g-}factor and spin of a single Mn ion, and $B_S$ is the Brillouin function. $T_0$ is the correction to Mn spin temperature and $x_{\text{eff}}$ is the effective concentration of Mn ions both accounting for the antiferromagnetic interaction between nearest neighbor Mn ions.\cite{gaj10,klo13} We introduce scaling factors $\gamma_e$ and $\gamma_h$, both smaller than unity, which account for the reduced overlap of the electron and hole wave functions with the \CdMnTe\ dot.\cite{mak00,hun04}

The corresponding magnetic field dependence of the transition linewidth $\Gamma(B)$ is is derived by summing contributions from $N^{e,h}_{\text{Mn}}$ Mn ions interacting with the electron and the hole, respectively. Obviously, in general $N^e_{\text{Mn}} \neq N^h_{\text{Mn}}$. The calculation is done in a muffin-tin approximation,\cite{klo13} i.e., we take a constant value of the wave function inside the dot volume and zero outside. The volumes for electron and hole wave functions are assumed equal to the \CdMnTe\ QD volume equal to the CdTe QD volume $V$. Within this approach, the number of interacting Mn ions can be calculated as $N^{e,h}_{\text{Mn}} = x_{\text{eff}} N_0 V^{e,h}_P$, where $V^{e,h}_P$ is, respectively, the volume of the electron and hole wave functions penetrating into the \CdMnTe\ dot. Thus, the scaling factors introduced above are given by: $\gamma_{e,h} = V^{e,h}_P/V$ and the expression for $\Gamma(B)$ reads:

\begin{multline}
\Gamma (B) =  \\
\sqrt{8 \ln 2 \cdot \left( \frac{\gamma_e \alpha^2-\frac{3}{2} \gamma_h \beta^2}{4V^2} \right) \frac{k_B T}{g \mu_B} N_{\text{Mn}} S \left( -\frac{\partial B_S}{\partial B}\right)}\, ,
\label{gammab}
\end{multline}
where $N_{\text{Mn}}$ is the total number of Mn ions in the \CdMnTe\ dot.

The lines shown in Fig. \ref{unc} are results of fitting of Eqs. \ref{eneb} and \ref{gammab} to the experimental data. The obtained fitting parameters are $E(0) = 1935.9$ meV, $x_{\text{eff}} = 0.048$, $T= 2$ K, $N_{\text{Mn}} =$ 120 and the CdTe QD volume expressed in the number of cation sites $N_0 V = $ 1940. This volume corresponds to a lens-shape dot roughly 2 nm high with a base diameter of 16 nm, which is a typical size for these QDs as shown by atomic force microscopy.\cite{woj08} The obtained number of Mn ions remains in a good agreement with a value expected for a QD with an intentional Mn concentration of 0.05.\cite{klo13}

For assessing the nature of the inter-dot coupling, the most important result is that we reproduce the experimental data with $\gamma_e = 0$ and $\gamma_h = 0.07$ for the $\smn$ polarization and $\gamma_h = 0.09$ for the $\spl$ polarization. Thus, we conclude that the observed Zeeman shifts and transition linewidths originate from a small penetration of the hole wave function from the CdTe dot into the \CdMnTe\ dot. However, this penetration is rather small, pointing to a weak tunnel coupling of the holes. Crucially, there is no resonant behavior, i.e. no change of $\gamma_h$ as a function of the magnetic field. A small difference between $\gamma_h$ for $\spl$- and $\smn$-polarized transitions points out that the hole states are far from resonance and that the field affects the wave function very weakly. However, the penetration is strong enough to induce efficient exciton spin relaxation and a strong polarization of the \Xz\ transition.

Although we are unable to perform analogous quantitative analysis for the broad transitions seen in Fig. \ref{maps}(b), we assume that the same mechanism governs their behavior. In this case, the exciton is predominantly localized in the \CdMnTe\ dot, with a part of the wave function leaking out and resulting in a Zeeman shift reduced with respect to an isolated \CdMnTe\ dot. Also, we remark that although the broad transitions and the narrow \Xm\ transition coincide at zero field, no coupling signatures are observed. We therefore conclude that the narrow and broad transitions seen in Fig. \ref{maps}(b) originate from two QDs located at a distance precluding their tunnel coupling.

\begin{figure}[!h]
  \includegraphics[angle=0,width=.33\textwidth]{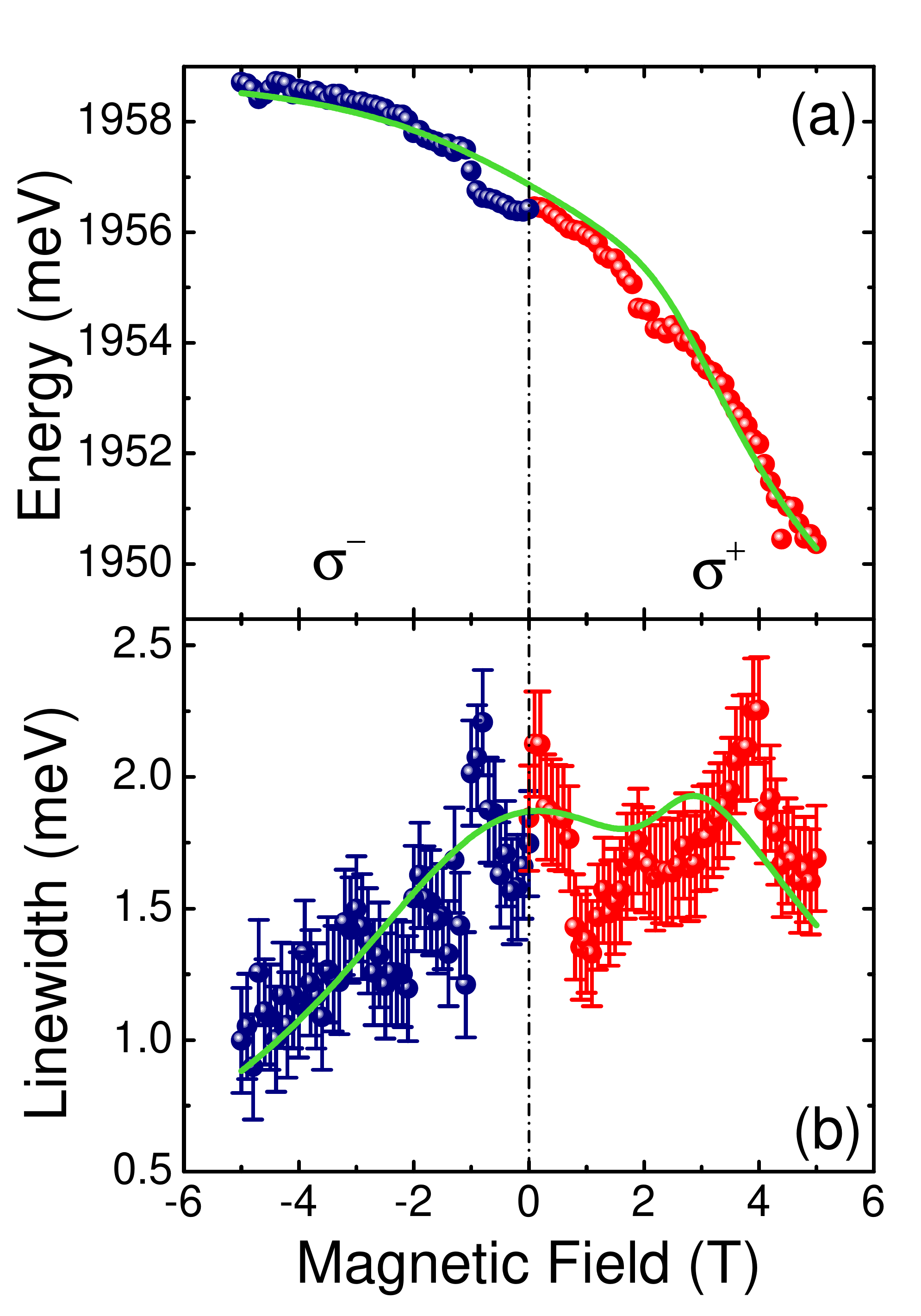}
  \caption{Energy positions (a) and linewidths (b) extracted from the PL spectra of a QDM presented in Fig. \ref{maps}(c). Lines are results of fitting of the theoretical model taking into account the formation of the molecular state. See text for details.}
  \label{coup}
\end{figure}

We now turn to the discussion of results presented in Fig. \ref{maps}(c). The PL dependence on magnetic field exhibits a complicated pattern. We single out the transition appearing at the highest energy -- 1956.4 meV, and interpret it as the recombination of the \Xz. Further below, we argue that the transitions at lower energies, exhibiting a complicated fine structure, correspond to the recombination of the \Xm. It is clear from the inspection of Fig. \ref{maps}(c) that the \Xz\ transition behaves differently in $\spl$ and in $\smn$ polarizations. This is shown in Fig. \ref{coup}, where we plot the field dependence of the transition energies and linewidths. The Zeeman shifts in $\spl$ and in $\smn$ polarizations amount in this case to 6.1 meV and 2.3 meV respectively. Also, the magnetic field dependence clearly deviates from the Brillouin function. Moreover, the transition linewidths for $\smn$ polarization are generally smaller than for $\spl$ polarization. In the former case, the linewidths decrease monotonically with the magnetic field, while in the latter a non-monotonic dependence is observed with a clear maximum at about 4 T.

The behavior of the transitions energies and linewidths seen in Fig. \ref{coup} is thus strikingly different from what is observed in Fig. \ref{unc}. In the present case, it is impossible to account for the experimental data assuming a field-independent wave function overlap with the Mn ions. It is clear that for the states participating in the recombination in $\spl$ polarization, the carrier-Mn ion exchange interaction becomes stronger with increasing magnetic field. As a result, both the Zeeman shift and the transition linewidths are larger than for the $\smn$ polarization.

In the following, we show that the transition behavior observed in Fig. \ref{maps}(c) and Fig. \ref{coup} is an optical signature of a spin-dependent coupling in a semimagnetic QDM. To reproduce the experimental data, we calculate the transition energies for $\spl$ and $\smn$ polarizations as $E_{\pm}(B) = E(0)+ \Delta E^{\pm}_h + \Delta E^{\mp}_e$, where $\Delta E_{e,h}$ are the exchange-induced energy shifts of the electron and hole states and the superscripts $\pm$ denote spin-up and spin-down states, respectively. Consistently with the interpretation of the data in Fig. \ref{unc}, for the hole, we assume a field independent wave function overlap with the \CdMnTe\ QD. Thus, for the hole states, we obtain (see Eq. \ref{eneb}): $\Delta E^{\pm}_{h} = \pm \frac{1}{2} \gamma_h N_0 \beta x_{\text{eff}} S B_{S}(\frac{g \mu B S}{k_B(T+T_0)})$. Electron energies $\Delta E^{\mp}_e$ are obtained by diagonalizing a simple Hamiltonian written in the basis of electron states in the CdTe and \CdMnTe\ QDs:\cite{sti06}
\beq
\hat{H}_{e} =
\left(
\begin{array}{cc}
0 & -t_e\\
-t_e & \Delta + E_e
\end{array}
\right)
\label{hame}
\eeq
where $t_e$ is the electron tunneling rate, $\Delta$ is the detuning between the electron states in zero field and $E_e = \pm \frac{1}{2} N_0 \alpha x_{\text{eff}} S B_{S}(\frac{g \mu B S}{k_B(T+T_0)})$ is the energy of the spin-up and spin-down electron states in the \CdMnTe\ dot. Here again we neglect the intrinsic Zeeman energy of the electrons in the CdTe QD.
We fit the measured transition energy dependence on the magnetic field assuming that the hole recombines with the electron in its lower-energy state, i.e., the bonding state. The results are presented as a line in Fig. \ref{coup}(a). Fitting parameters are: $\Delta = 121$ meV, $t_e = 1.6$ meV, $x_{\text{eff}}=0.075$, $T=4.4$ K, and the field independent hole overlap $\gamma_h$ = 0.023. The agreement with the experimentally measured values is very good. To calculate the magnetic field dependence of the transition linewidths, we apply the formula given in Eq. \ref{gammab} with $\gamma_e$ being the magnetic field dependent probability of finding the electron in the \CdMnTe\ QD. Thus, $\gamma_e$ is calculated as a projection of the eigenvector of the Hamiltonian (\ref{hame}) onto the \CdMnTe\ QD electron state. The fitting is performed by keeping the parameters obtained in fitting of the transition energies and adjusting only the number of the Mn ions $N_{\text{Mn}}$ and the volume of the CdTe dot $V$. Results are presented as a line in Fig. \ref{coup}(b). We obtain $N_{\text{Mn}}$ = 170 and $N_0 V$ = 2000. Remarkably, the fitted curve reproduces the all the features of the linewidth dependence on magnetic field: narrowing of the transition in $\smn$ polarization, the initial decrease in $\spl$ polarization and finally the maximum at about 4 T.

\begin{figure}[!h]
  \includegraphics[angle=0,width=.33\textwidth]{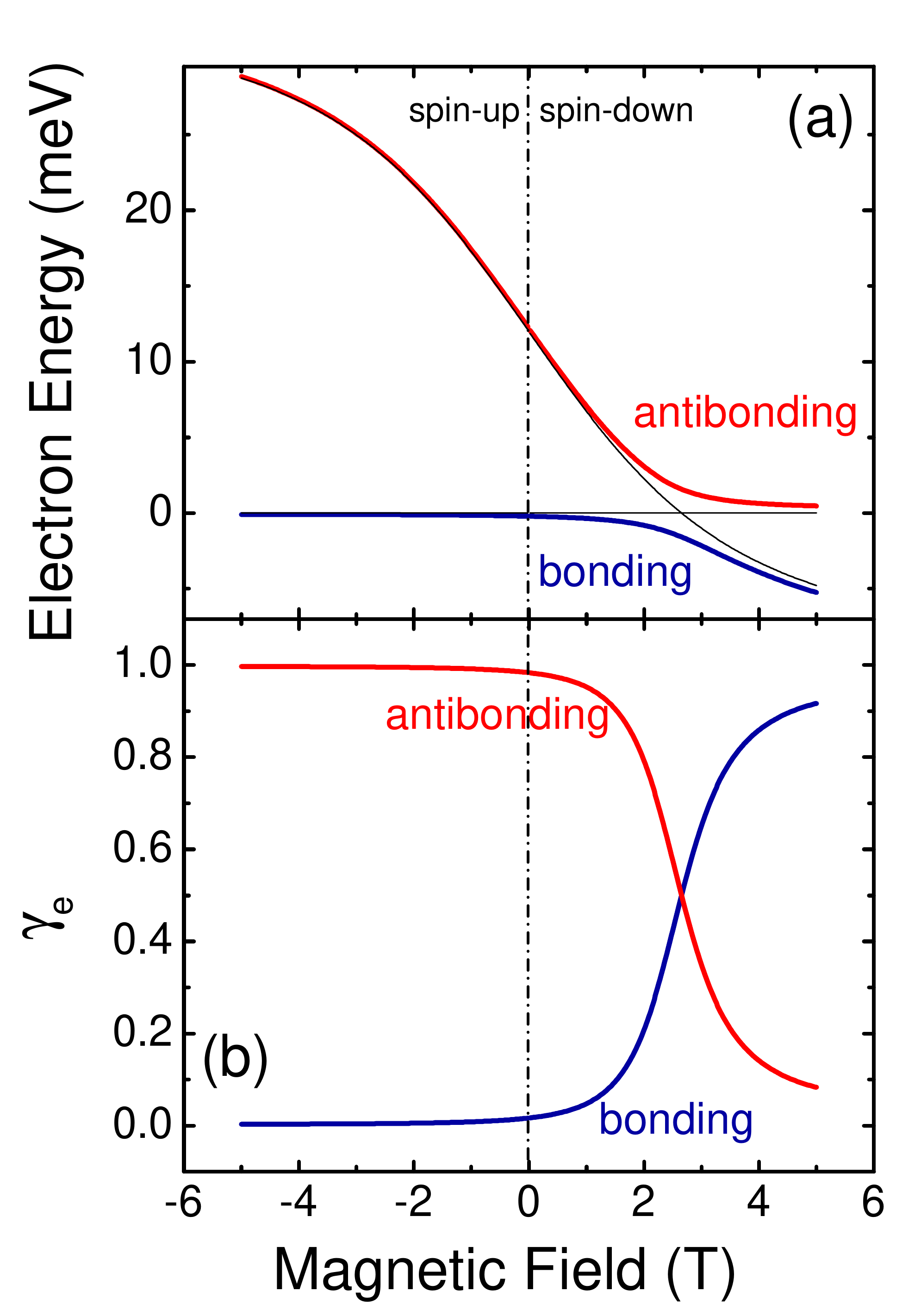}
  \caption{(a) Thick lines: electron energies vs. magnetic field evaluated in fitting of the experimental data in Fig. \ref{coup}. The avoided crossing at about 3 T demonstrates the formation of a molecular state for the spin-down electrons. Thin lines: energies of uncoupled electronic states. (b) Probability of finding an electron in the \CdMnTe\ QD for the two lowest electron orbitals of the QDM.}
  \label{model}
\end{figure}

Electron energies obtained to fit the results in Fig. \ref{coup} are plotted as a function of the magnetic field in Fig. \ref{model}(a). The figure demonstrates the spin-specific nature of the molecular coupling in the QDM. The molecular state is formed at the anticrossing of the electron levels at about 3 T, where the spin-down states come into resonance. No resonance is reached for the spin-up electrons. In fact, these states, which contribute to the emission in $\smn$ polarization, are almost purely localized in either CdTe or \CdMnTe\ dot. This effect is further demonstrated in Fig. \ref{model}(b), where we plot $\gamma_e$ as a function of the magnetic field. For spin-up states, we obtain $\gamma_e \ \approx 0$ and $\gamma_e \ \approx 1$ for the low and high energy electron states, respectively. Conversely, for spin-down electrons the low energy, bonding state progressively shifts from being predominantly localized in the CdTe dot at small magnetic fields, through a complete hybridization at about 3 T, to being localized mostly in the \CdMnTe\ QD at high fields.

\section{Discussion and Conclusions}

It is important to clarify the difference in coupling regimes discussed above. For the QD studied in Fig. \ref{maps}(b) and Fig. \ref{unc}, we deal with a weak and nonresonant coupling. The wave function penetrates from the CdTe to \CdMnTe\ QD (or a \CdMnTe\ wetting layer) giving rise to an increased (with respect to a CdTe dot) Zeeman splitting and linewidth.
A qualitatively different type of coupling is present in the QDM investigated in Fig. \ref{maps}(c) and Fig. \ref{coup}. While the penetration of the hole wave function is constant with the field, the electron coupling is spin-selective. The wave functions of spin-up states remain unaffected by the magnetic field. On the contrary, the spin-down states states are brought to resonance which results in strong redistributions of the wave functions.
We remark that an intermediate coupling regime is possible and indeed was reported for pairs of CdTe and \CdMnTe\ dots fabricated using a selective interdiffusion technique from CdTe quantum wells.\cite{bac08} In that case, the penetration of the wave function does change with the magnetic field, but the resonance is not reached and the molecular state is not formed.

The signatures of molecular coupling between our QDs significantly differ from those observed in non-magnetic systems. As shown above, the main difference is the spin-specific nature of the coupling. The other important difference stems from efficient exciton spin and energy relaxation, which drives the system of the exciton and the Mn ions to thermal equilibrium. The relaxation occurs as a result of strong exchange interaction between the exciton and the Mn ions. As shown in Ref. [\cite{klo11pol}], the exciton spin relaxation rate strongly increases with the typical effective field exerted by Mn ions. This effect is also seen in the results presented here. Comparing the polarization degrees of the \Xz\ transitions in Figs. \ref{maps}(b) and (c), we conclude that a more efficient relaxation occurs in the former case. This can be understood by examining the exciton states recombining in $\smn$ polarization -- the ones undergoing the relaxation to lower energy. As shown by fitting of the data in Figs. \ref{unc} and \ref{coup} for these states, the penetration of the wave function into the \CdMnTe\ dot is larger in the former case, hence the faster spin relaxation.

Another manifestation of exciton energy relaxation in the \CdMnTe\ QD is the absence of PL signal from the anti-bonding state. Indeed, since the tunnel coupling results in a splitting of bonding and anti-bonding states by more than 2 meV (see Fig. \ref{model}(a)), the relative occupation of the higher and lower energy states in thermal equilibrium becomes larger than 10\% only for exciton temperatures higher than 12 K. We expect that due to strong exciton-Mn ion interaction, the exciton system is in equilibrium with the Mn spin system and thus remains at a much lower temperature.

The coupling signatures related to the \Xm\ transition require further studies and a separate analysis. We note that in the case of a QDM, the initial state of the recombination comprises of two electrons and a hole. Electron tunneling between the QDs results in six configurations of the initial state, related to the six spin configurations of the electron pair.\cite{sti06,sch07prb} The final state is a single electron in either bonding or anti-bonding state. Thus, one deals with 12 possible transitions in each light polarization. Energy relaxation towards thermal equilibrium results in occupation of only the lowest of the initial states, but even if a single state is occupied this still results in two transitions in each polarization. Indeed, as seen in Fig. \ref{maps}(c), a multitude of transitions is observed for the \Xm\ recombination from the QDM. Since a significant part of the hole confinement results from Coulomb attraction to the electron, we speculate that the form of the hole wave function may strongly depend on the two electron configuration. In other words,
a proper analysis of the charged exciton state in a semimagnetic QDM requires taking into account electron-hole Coulomb correlations. Also, as shown in the studies of the coupled CdSe and CdZnMnSe QDs, a specific spin-dependent coupling mechanism, leading to  antiferromagnetic coupling of electron spin,\cite{lee05} needs to be considered.

In conclusion, we have presented evidence for controllable coupling in a semimagnetic quantum dot molecule comprising of a CdTe and a \CdMnTe\ dot. Giant Zeeman effect resulting from the exchange interaction between the carriers and Mn ions allows to tune electron states to resonance. Tunnel coupling results in a formation of a molecular state overlapping both dots. The spectroscopic signature of the coupling is the resonant enhancements of the Zeeman shift and transition linewidth both accounted for in a theoretical model. Our results show that the proposed electrical tuning of the electron {\em g-}factors should indeed be possible when the quantum dot molecule is placed in a field-effect structure. Furthermore, these nanostructures provide an excellent platform for studies of spin dependent tailoring of carrier wave functions -- a crucial task for developing efficient quantum gates for information processing.

This research was supported by the Polish National Science Center grant no. 2011/01/B/ST3/02287. T.W. acknowledges also the support from the Foundation for Polish Science through the International Outgoing Scholarship 2014.

\end{document}